\documentclass[aps,twocolumn,prl,superscriptaddress]{revtex4-1}
\usepackage{natbib}
\usepackage{graphicx}
\usepackage{amsmath}
\usepackage{amssymb}

\def\ket#1{{\left| #1 \right\rangle}}

\begin{document}

\title{Correspondence: Enhancing a phase measurement by sequentially probing a solid state system}
\author{P. A. Knott}
	\email{P.Knott@Sussex.ac.uk}
	\affiliation{Department of Physics and Astronomy, University of Sussex, Brighton BN1 9QH, United Kingdom}
\author{W. J. Munro}
	\affiliation{NTT Basic Research Laboratories, NTT Corporation, 3-1 Morinosato-Wakamiya, Atsugi, Kanagawa 243-0198, Japan}
\author{J. A. Dunningham}
	\affiliation{Department of Physics and Astronomy, University of Sussex, Brighton BN1 9QH, United Kingdom}


\date{\today}

\maketitle



In a recent paper, Liu \textit{et. al.} \cite{liu2015demonstration} claim to perform the first room temperature entanglement-enhanced phase measurement in a solid-state system. We argue here that this claim is incorrect: their measurement is not enhanced because of the entanglement in their system, but instead the enhancement comes from the fact that the phase shift is applied twice to their state.

Typically a quantum metrology experiment involves three key stages: a) state preparation, b) phase shift induced by an external system, and c) measurement/read-out \cite{giovannetti2006quantum,giovannetti2011advances,demkowicz2012elusive}. Normally, the phase shift $\phi$ is imprinted on the system during stage b) by a unitary operator $\hat{U}(\phi)$, and if an entangled state has been prepared then the measurement of the phase shift can be entanglement-enhanced \cite{giovannetti2006quantum,giovannetti2011advances,demkowicz2012elusive,escher2011general}. However, in \cite{liu2015demonstration} the phase shift is applied during the state preparation stage, and as a result this scheme is not able to measure an external parameter, as is usually the case in quantum metrology. Furthermore, in \cite{liu2015demonstration} the phase shift is applied twice, and one phase shift is applied when there is no entanglement in the system. 

In \cite{liu2015demonstration} the authors start with the state $\ket{0\uparrow}$ where $\ket{\uparrow}$ ($\ket{\downarrow}$) is the nuclear spin up (down), and the electron spin can be in the ground state $|0\rangle$ or the excited state $|1\rangle$. Using a radio frequency (RF) pulse this state evolves to $|\Psi\rangle = {1 \over \sqrt{2}} ( \ket{0\uparrow}+e^{i \phi} \ket{0\downarrow} )$. Next (and later in time) a microwave (MW) pulse transforms the state into $ |\Psi\rangle = {1 \over \sqrt{2}} ( \ket{1\uparrow}+e^{2i \phi} \ket{0\downarrow})$. The relative phase shift is now $2\phi$, which leads to the $\sqrt{2}$ enhancement in the phase uncertainty demonstrated in the paper. The question we raise here, which is critical to the authors' claim, is whether the origin of this enhancement can be attributed to entanglement or not.

We claim that the authors obtain an enhancement because they have multiple applications of the phase shift at different times. This strategy has been demonstrated by Higgins \textit{et. al.} \cite{higgins2007entanglement} which works as follows: We start with a single spin in a superposition state $\ket{\Psi} = {1 \over \sqrt{2}} ( |0\rangle+ |1\rangle)$. We can then apply a phase shift, $\hat{U}(\phi) = e^{i \hat{n} \phi}$ where $\hat{n}$ is the number operator which counts the number of spins in the excited state, giving
\begin{equation}
\ket{\Psi} = {1 \over \sqrt{2}} ( |0\rangle+ |1\rangle)  \rightarrow {1 \over \sqrt{2}} ( |0\rangle+ e^{i \phi} |1\rangle).
\end{equation}
We can then apply $\hat{U}(\phi)$ for a second time, for example if measuring a magnetic field we subject the spin to the magnetic field for a second time, to get:
\begin{equation}
e^{i \hat{n} \phi} \left( {1 \over \sqrt{2}} ( |0\rangle+ e^{i \phi} |1\rangle) \right) = {1 \over \sqrt{2}} ( |0\rangle+ e^{2i \phi} |1\rangle ) 
\end{equation}
We now have the same phase shift as in \cite{liu2015demonstration}, using a similar method (i.e. applying the phase shift twice), but without entanglement.

We now look at the usual method for using entanglement to enhance a phase measurement in a spin system \cite{huelga1997improvement,schaffry2010quantum,leibfried2004toward,matsuzaki2011magnetic}. We start with a Bell state of $2$ spins:
\begin{equation}
|\Phi^+ \rangle = {1 \over \sqrt{2}} ( |0 0\rangle + |1 1\rangle )
\end{equation}
Applying the phase shift operator {\it only once} to this state gives
\begin{equation}
{1 \over \sqrt{2}} ( |0 0\rangle + e^{i 2 \phi} |1 1\rangle).
\end{equation}
Here the $2$-fold enhancement in the phase shift is directly caused by the entanglement: $2$ spins are entangled, and therefore the phase shift has a factor of $2$. This is not the case in \cite{liu2015demonstration}: the factor of $2$ that is observed comes from the fact that the authors have applied a phase shift twice -- once using RF and once using MW. Furthermore, in \cite{huelga1997improvement} it is shown that for a two-level system in the presence of dephasing (as will be the case in \cite{liu2015demonstration}), maximally entangled states cannot perform better than uncorrelated states, and the apparent contradiction between \cite{liu2015demonstration} and these results should be addressed.

We now take a closer look at the system in \cite{liu2015demonstration}. The RF $\pi/2$ pulse has the following effect:
\begin{align}
|0\uparrow\rangle \xrightarrow{\text{RF}} {1 \over \sqrt{2}} ( \ket{0\uparrow}+e^{i \phi_{_{\text{RF}}}} \ket{0\downarrow}).
\end{align}
Here, unlike in \cite{liu2015demonstration}, we label the phase shift as $\phi_{_{\text{RF}}}$. This phase shift has been applied without any entanglement in the system. The authors then apply a selective MW $\pi$ pulse for the transition $\ket{0\uparrow} \leftrightarrow \ket{1\uparrow}$. This induces a phase $\phi_{_{\text{MW}}}$, and therefore the effect on the whole state is
\begin{align}
{1 \over \sqrt{2}} (\ket{0\uparrow}&+e^{i \phi_{_{\text{RF}}}} \ket{0\downarrow}) \\ &\xrightarrow{\text{MW}} {1 \over \sqrt{2}} ( \ket{1\uparrow}+e^{i (\phi_{_{\text{RF}}}+\phi_{_{\text{MW}}})} \ket{0\downarrow})\notag.
\end{align}
The authors are able to control the relative phase shifts of the RF and MW pulses, enabling them to set $\phi=\phi_{_{\text{RF}}}=\phi_{_{\text{MW}}}$. Thus the final state is ${1 \over \sqrt{2}} (\ket{1\uparrow}+e^{i 2\phi} \ket{0\downarrow})$. A bipartite entangled state has been created, with a relative phase of $2\phi$ between the two parts of the superposition, and the authors attribute the factor of $2$ to the entanglement. Note that if $\phi_{_{\text{RF}}} \neq \phi_{_{\text{MW}}}$ the illusion of an entanglement-enhanced phase shift would be broken, and furthermore, if we set $\phi_{_{\text{RF}}} = - \phi_{_{\text{MW}}}$ then there would be no relative phase shift at all.

A sequential application of phases can sometimes mimic entanglement-enhanced measurements, but is not the same thing. In \cite{liu2015demonstration}, where the phases are applied to different qubits, entanglement is needed to \textit{enable} the sequential application of the phases but, importantly, the entanglement is just a detail of the particular experiment. It is not what enhances the measurement.

We have seen that the $2\phi$ phase shift in \cite{liu2015demonstration} arises due to the state preparation method; this same phase shift can also be created with no entanglement present in the system. If the authors replace their $\pi/2$ RF pulse with two $\pi/4$ RF pulses, we would find
\begin{align}
\ket{0\uparrow} \xrightarrow{} {1 \over \sqrt{2}} (\ket{0\uparrow}+e^{i 2\phi} \ket{0\downarrow}).
\end{align} 
Alternatively, a similar entangled state to that which the authors create can be made, at least in principle, by applying a single pulse with a frequency of RF + MW. This would allow the following transformation:
\begin{align}
\ket{0\uparrow} \xrightarrow{\text{RF+MW}} {1 \over \sqrt{2}} (\ket{0\uparrow}+e^{i \phi} \ket{1\downarrow}).
\end{align}
These examples show that, if the phase shift is applied in the preparation state, then we can effectively have any phase shift we want, and simple alterations to the scheme in \cite{liu2015demonstration} can produce an enhancement without entanglement, or an entangled state with no phase at all. The entanglement is therefore not the necessary part of the enhancement in \cite{liu2015demonstration}.

In conclusion, the experiment performed in \cite{liu2015demonstration} shows impressive control and manipulation of a solid state system. However, we have argued here that the phase-estimation performed is not ``entanglement-enhanced'' in the usual sense and can instead be attributed to the sequential application of the phase shift during state preparation. \\

\subsection{Acknowledgements}

This work was partly funded by the UK EPSRC through the Quantum Technology Hub: Networked Quantum Information Technology (grant reference EP/M013243/1). \\


\subsection{Author contributions}

P.A.K. conceived the main argument of the criticism. W.J.M. and J.A.D. helped develop the argument. All authors reviewed the manuscript.


\subsection{References}
\bibliographystyle{unsrt}
\bibliography{MyLibrary_NV_comment_Sept2015}

\end{document}